\documentstyle[preprint,prb,aps,psfig]{revtex}
\begin{document}

\title{Irreversible Bimolecular Reactions of Langevin Particles}

\author{D. J. Bicout$^{1}$\,, A. M. Berezhkovskii\thanks{ 
Permanent address: Karpov Institute of Physical Chemistry, Ul.
Vorontsovo Pole 10, 103064, Moscow K-64, Russia}$^{,2}$\, and  Attila Szabo$^2$}

\address{$^{1}$ Institut Laue-Langevin, Theory Group \\ and 
INFM-Operative Group Grenoble CRG IN13, \\ 
B. P. 156, 38042 Grenoble Cedex 9, France.\\
$^{2}$ Laboratory of Chemical Physics, 
National Institute of Diabetes and Digestive and Kidney\\ Diseases, 
National Institutes of Health, Bethesda, Maryland 20892}

\maketitle

\begin{abstract}
The reaction $A+B\,\longrightarrow\,B$ is studied when the reactants 
diffuse in phase space, i.e. their dynamics is described by the 
Langevin equation. The steady-state rate constants are calculated for 
both the target (static $A$ and mobile $B$'s) and trapping (mobile $A$ and 
static $B$'s) problems when the reaction is assumed to occur at the first 
contact. For Brownian dynamics (i.e., ordinary diffusion), the rate 
constant for both problems is a monotonically decreasing function of the 
friction coefficient $\gamma$. For Langevin dynamics, however, we find 
that the steady-state rate constant exhibits a turnover behavior as a 
function of $\gamma$ for the trapping problem but not for the target 
problem. This turnover is different from the familiar Kramers' turnover 
of the rate constant for escape from a deep potential well because the 
reaction considered here is an activationless process. 
\end{abstract}



\newpage
\baselineskip 0.75cm

\renewcommand{\theequation}{\arabic{section}.\arabic{equation}}
\newcounter{xeq}

\section{Introduction}
\setcounter{equation}{0}

Theories of bimolecular chemical reactions in solvents that occur at first 
contact of the reactants has been traditionally developed assuming that 
molecules move in diffusive manner \cite{Rice}. This corresponds 
to the so-called high friction limit of the more general Langevin dynamics. 
In this paper we develop a theory for Langevin reactant dynamics and 
analyze the kinetics over the entire range of the friction coefficient, 
from low to high friction.

Specifically, we study the kinetics of the irreversible reaction 
$A+B\,\longrightarrow\,B$ in two limiting cases: static $A$ and mobile $B$'s 
(target problem) and mobile $A$ and static $B$'s (trapping problem); the 
difference between the two problems is illustrated in Fig. 1. 
Traditional theories based on the diffusion equation predict monotonic 
decrease of the reaction rate when the diffusion constant decreases, i.e. 
when the friction increases. It will be shown that in the case of the 
Langevin dynamics such a  behavior of the reaction rate with the friction 
is observed only for the target problem. For the trapping problem the 
steady state rate constant has a turnover behavior as a function of the 
friction coefficient $\gamma$: it first increases with $\gamma$, reaches 
a maximum, turns over and then decreases approaching zero as 
$\gamma\rightarrow\infty$. This  rate constant vanishes at $\gamma=0$ 
in one dimension while it remains finite in higher dimensions.

The turnover behavior of the reaction rate is well known in the Kramers' 
theory of activated unimolecular reactions \cite{Kramers,Melnikov}. This theory 
considers the escape of a Langevin particle from a deep potential well. To 
escape the particle has to overcome a high (compared to ${\rm k_BT}$) 
potential barrier. The escape rate vanishes as $\gamma\rightarrow 0$ and 
as $\gamma\rightarrow\infty$ and has a maximum in between. At low friction 
the rate is limited by energy exchange with the environment and vanishes at 
$\gamma=0$ since the particle cannot gain the energy necessary to 
overcome the barrier. As $\gamma$ becomes finite, the particle can  
exchange energy with the environment. This leads to the increase of 
the reaction rate. This is the so-called low friction or energy diffusion 
regime. In the opposite limiting case when $\gamma\rightarrow\infty$ 
(the high friction regime), particle's motion is purely diffusive. As a 
consequence, the escape rate is proportional to the diffusion constant, 
$D={\rm k_BT}/\gamma$ so that the reaction rate vanishes when $D$ goes to 
zero. 

In the trapping problem a particle does not need to overcome an energy 
barrier in order to react. In fact, even when the friction is zero 
every particle with finite initial velocity will eventually react. Of 
course, particles that move faster will be trapped more quickly and 
are the first to be removed from the system. When the friction  
increases from zero, these particles can be replaced by initially 
slower moving particles that gain energy from the heat bath. Thus 
the trapping rate at first increases as the friction increases. As 
the friction is further increased, the motion of the particle eventually 
becomes diffusive and in this limit the trapping rate is proportional 
to the diffusion coefficient $D$ which vanishes as the friction tends 
to infinity.  

To analyze the kinetics when motion of the reactants is governed by the 
Langevin equation one has to solve the Klein-Kramers equation 
in the phase space with appropriate boundary conditions. This is 
an extremely complicated task. Even the simple problem of survival of a 
free particle moving on a line in the presence of a single trap  has 
an extremely complicated solution in the phase space. This solution was 
found by Marshall and Watson \cite{Marsh} forty years later after the 
problem was posed by Wang and Uhlenbeck \cite{Wang} in $1945$. A 
discussion of the source of this complexity and a list of key references 
can be found in Refs.[\onlinecite{Marsh,Doering}].

Since we cannot solve the problem in phase space exactly we will find 
approximate solutions in the limiting cases of low and high friction. 
These solutions are then used to construct an interpolation formula 
that covers the entire range of the friction coefficient. The 
outline of the paper is as follows. In section II we analyze the 
survival of a Langevin particle moving on a line between two static 
traps. Some of results derived in this section has been reported 
earlier \cite{Bicout} without derivation. An approximate solution 
found in Section II is used in Section III to calculate the 
steady-state rate constant for the trapping problem in one 
dimension over the entire range of the friction coefficient. Since 
this approach is not generalizable to higher dimensions, in Section IV 
we develop another approximate method that can be applied in a space 
of arbitrary dimension and use it in Section V to find the 
rate constants for the target and trapping problems as a function 
of the friction coefficient in one and three dimensions. Section VI 
contains a brief summary.

\section{Survival on an Interval Terminated by Trapping Points}
\label{sec:Brow}
\setcounter{equation}{0}

In this section we analyze the survival of a Langevin particle moving on a 
line between two traps. We calculate the effective rate constant 
characterizing the kinetics of trapping. By averaging this rate with respect 
to the length of the interval, we will obtain the steady-state rate constant 
for the trapping problem in one dimension over the entire range of the 
friction. 

Let $x$ and $\dot{x}=v$ denote the position and velocity of a free  
particle of unit mass whose dynamics is governed by the Langevin equation,
\begin{equation}
\ddot{x}+\gamma\,\dot{x}=R(t)\:,
\label{lang}
\end{equation}
where $\gamma$ is the friction coefficient and $R(t)$ is a Gaussian random 
force of zero mean with correlation function given by the 
fluctuation-dissipation relation, 
$\langle R(t)R(t')\rangle=2\,\gamma\,\delta(t-t')$. Here and 
below the thermal energy is set ${\rm k_BT}=1$. 

Equivalently, the joint probability density, $P(x,v,t)$, of finding the 
particle at the phase space point $(x,v)$ at time $t$ is described by the 
Klein-Kramers equation:
\begin{equation}
\frac{\partial P(x,v,t)}{\partial t}=\left[-v\,
\frac{\partial}{\partial x}\,+\,\gamma\,
\frac{\partial}{\partial v}\,{\rm e}^{-v^2/2}\,
\frac{\partial}{\partial v}\,{\rm e}^{v^2/2}\right]\,P(x,v,t)\:.
\label{KK}
\end{equation}
To study the escape of the Brownian particle from the interval $[0,1]$, 
we require that $P(x,v,t)$ satisfies the absorbing boundary conditions:
\begin{eqnarray}
\left\{\begin{array}{lcc}
P(x,v>0,t)=0 & \mbox{at} & x=0\:,\\
P(x,v<0,t)=0 & \mbox{at} & x=1\:.
\end{array}\right.
\label{bc}
\end{eqnarray}
Assuming that the system has initially a uniform spatial distribution 
and the Maxwell velocity distribution, the survival probability that 
describes the fate of the particle in the interval is given by,
\begin{equation}
S(t)=\int_{0}^{1}\!dx\int_{-\infty}^{+\infty}\!dv\,P(x,v,t)\:,
\label{sur}
\end{equation}
where $P(x,v,t)$ is the solution of Eq.(\ref{KK}) with initial condition 
$P(x,v,t=0)=p_{\rm eq}(x)\,f_{\rm eq}(v)$, with $p_{\rm eq}(x)=1$ and
\begin{equation} 
f_{\rm eq}(v)=\frac{{\rm e}^{-v^2/2}}{\sqrt{2\pi}}\:.
\label{feq}
\end{equation}
To characterize the kinetics of the escape from the interval we will 
use the effective rate constant, $k$, defined as the reciprocal of the 
mean lifetime of the particle on the interval,
\begin{equation}
\frac{1}{k}=\int_{0}^{\infty}\!S(t)\,dt\:.
\label{k}
\end{equation} 

When $\gamma=0$, the particle moves with a constant velocity. The survival 
probability of a particle starting at $x$ with a velocity $v$ is,
\begin{equation}
S(t|x,v)={\rm H}(-v)\,{\rm H}\left[x+vt\right]+
{\rm H}(v)\,{\rm H}\left[x-1-vt\right]\:,
\label{soft0}
\end{equation}
where ${\rm H}(z)$ is the Heaviside step function defined as ${\rm H}(z)=0$ 
for $z<0$ and ${\rm H}(z)=1$ for $z>0$. Averaging $S(t|x,v)$ over the uniform 
distribution in $x$ and the equilibrium distribution in $v$ gives:
\begin{eqnarray}
S(t)=\int_{0}^{1}\!dx\int_{-\infty}^{+\infty}\!S(t|x,v)\,f_{\rm eq}(v)\,dv 
={\rm erf}\left[\frac{1}{\sqrt{2}\:t}\right]\,
-\,\sqrt{\frac{2}{\pi}}\:t\,\left[1-{\rm e}^{-1/(2t^2)}\right]\:.
\label{sexact}
\end{eqnarray}
This expression shows that in the ballistic limit, $S(t)$ has a power law 
tail  
\begin{equation}
S(t)\simeq \frac{1}{\sqrt{2\pi}\:t}
\:\:\:;\:\:\:t\rightarrow \infty\:. 
\label{sinf}
\end{equation}
From this and the definition in Eq.(\ref{k}) it follows that $k=0$ when 
$\gamma=0$. 

On the other hand, when $\gamma=\infty$, the particle does not move and 
$S(t)=1$. As a consequence, $k=0$ in this limiting case also. Since $k$ 
vanishes in the $\gamma\rightarrow 0$ and $\gamma\rightarrow \infty$ 
limits, this rate constant shows a turnover behavior as a function of $\gamma$.

Since the solution of Eq.(\ref{KK}) with boundary conditions (\ref{bc}) 
is unknown, we adopt the following strategy for determining $k$ for arbitrary 
values of $\gamma$. We first derive approximate expressions for $k$ in high 
and low friction regimes, and then use the  
Visscher-Mel'nikov-Meshkov~\cite{Melnikov,Visscher} (VMM) interpolation 
formula to obtain an expression for the rate constant that covers the 
entire range of $\gamma$.

\subsection{High Friction Regime} 
\label{sec:Pdiff}

In this section, we analyze the the high friction limit. First, we 
reduce the Klein-Kramers equation (\ref{KK}) with absorbing boundary 
conditions (\ref{bc}) to the ordinary diffusion equation with radiation 
boundary conditions. Next, we calculate the effective rate constant by 
solving this equation. 

To proceed, we define the reduced probability density, 
\begin{equation}
p(x,t)=\int_{-\infty}^{+\infty}\!P(x,v,t)\,dv\:,
\end{equation}
and the flux, 
\begin{equation}
j(x,t)=\int_{-\infty}^{+\infty}\!v\,P(x,v,t)\,dv\:.
\label{j1}
\end{equation}
Integration of Eq.(\ref{KK}) over the velocity leads to the continuity 
equation:
\begin{equation}
\frac{\partial p(x,t)}{\partial t}=-\,
\frac{\partial j(x,t)}{\partial x}\:,
\label{KKpj}
\end{equation}
in which we have used $P(x,\pm\infty,t)=0$. Multiplying Eq.(\ref{KK}) by 
$v$ and integrating over the velocity from $-\infty$ to $+\infty$, we have:
\begin{equation}
\frac{\partial j(x,t)}{\partial t}=-\gamma\,j(x,t)\,-\,
\int_{-\infty}^{+\infty}\!v^2\,\frac{\partial P(x,v,t)}{\partial x}\,dv\:.
\label{jx}
\end{equation}
This equation can easily be solved to give:
\begin{equation}
j(x,t)={\rm e}^{-\gamma t}\,j(x,0)\,-\,
\int_{0}^{t}\!dt'\,{\rm e}^{-\gamma (t-t')}\,
\int_{-\infty}^{+\infty}\!v^2\,\frac{\partial P(x,v,t')}{\partial x}\,dv\:.
\end{equation}
Note that since the initial distribution of the velocity is Maxwellian, we 
have $j(x,0)=0$. In the $\gamma\gg 1$ limit, $P(x,v,t')$ does not vary on 
times  of order of $1/\gamma$. Therefore, the integration over $t'$ can be 
carried out easily to give:
\begin{equation}
j(x,t)=-\frac{1}{\gamma}\,
\int_{-\infty}^{+\infty}\!v^2\,
\frac{\partial P(x,v,t)}{\partial x}\,dv\:.
\label{j2}
\end{equation} 
Plugging this into Eq.(\ref{KKpj}) leads to:
\begin{equation}
\frac{\partial p(x,t)}{\partial t}=
\frac{1}{\gamma}\,
\int_{-\infty}^{\infty}\!v^2\,\frac{\partial^2 P(x,v,t)}{\partial x^2}\,dv\:.
\label{KKpp}
\end{equation}
In the $\gamma\gg 1$ limit, $P(x,v,t)$ can be written as an expansion in 
powers of $1/\gamma$. The zero order term of this expansion is, 
\begin{equation}
P(x,v,t)=f_{\rm eq}(v)\,p(x,t)\:.
\label{pxv}
\end{equation}
Using this in Eq.(\ref{KKpp}) one obtains the ordinary one-dimensional 
diffusion equation,
\begin{equation}
\frac{\partial p(x,t)}{\partial t}=D\,
\frac{\partial^2 p(x,t)}{\partial x^2}\:\:\:;\:\:\:D=\frac{1}{\gamma}\:.
\label{KKsd}
\end{equation}

To derive boundary conditions supplementing this equation we use the two 
expressions for the flux given in Eqs.(\ref{j1}) and (\ref{j2}) 
to write:
\begin{eqnarray}
j(x,t) & = & \int_{-\infty}^{0}\!v\,P(x,v,t)\,dv+\int_{0}^{\infty}\!v\,
P(x,v,t)\,dv \nonumber \\
& = & -\frac{1}{\gamma}\,
\int_{-\infty}^{0}\!v^2\,\frac{\partial P(x,v,t)}{\partial x}\,dv\,
-\,\frac{1}{\gamma}\,
\int_{0}^{\infty}\!v^2\,\frac{\partial P(x,v,t)}{\partial x}\,dv\:,
\label{j3}
\end{eqnarray}
where the integral over the velocity has been broken into two parts. At the 
boundaries, according to Eq.(\ref{bc}), $P(0,v>0,t)=P(1,v<0,t)=0$. 
This suggests that $\left.\partial P(x,v>0,t)/\partial x\right|_{x=0}$ 
is much larger than $\left.\partial P(x,v<0,t)/\partial x\right|_{x=0}$, and 
$\left.\partial P(x,v<0,t)/\partial x\right|_{x=1}$ is much larger 
than $\left.\partial P(x,v>0,t)/\partial x\right|_{x=1}$, when $|v|$ is not 
too small. Since contribution of small velocities are weakened by the 
multiplying factor $v^2$, we neglect these small terms. Retaining only 
major contributions leads to:
\setcounter{xeq}{\value{equation}}
\addtocounter{xeq}{1}
\renewcommand{\theequation}{\arabic{section}.\arabic{xeq}\alph{equation}}
\setcounter{equation}{0}
\begin{eqnarray}
j(0,t)=\left.\left(-\frac{1}{\gamma}\,
\int_{0}^{\infty}\!v^2\,\frac{\partial P(x,v,t)}{\partial x}
\,dv\right)\right|_{x=0} & = & 
\int_{-\infty}^{0}\!v\,P(0,v,t)\,dv\:, \label{bcpa} \\
j(1,t)=\left.\left(-\frac{1}{\gamma}\,
\int_{-\infty}^{0}\!v^2\,\frac{\partial P(x,v,t)}{\partial x}
\,dv\right)\right|_{x=1} & = & 
\int_{0}^{\infty}\!v\,P(1,v,t)\,dv\:.  \label{bcpb} 
\end{eqnarray}
\renewcommand{\theequation}{\arabic{section}.\arabic{equation}}
\setcounter{equation}{\value{xeq}}
Substituting the approximation for $P(x,v,t)$ given in Eq.(\ref{pxv}) into 
these relations, we obtain the radiation boundary conditions:
\setcounter{xeq}{\value{equation}}
\addtocounter{xeq}{1}
\renewcommand{\theequation}{\arabic{section}.\arabic{xeq}\alph{equation}}
\setcounter{equation}{0}
\begin{eqnarray}
\left.D\,\frac{\partial p(x,t)}{\partial x}\right|_{x=0}=\kappa\,\,p(0,t)\:,
\label{rbca} \\
\left.-D\,\frac{\partial p(x,t)}{\partial x}\right|_{x=1}=\kappa\,p(1,t)\:,
\label{rbcb}
\end{eqnarray}
\renewcommand{\theequation}{\arabic{section}.\arabic{equation}}
\setcounter{equation}{\value{xeq}}
where $\kappa$ is defined as,
\begin{equation}
\kappa=\langle |v|\rangle=
\int_{-\infty}^{\infty}\!|v|\,f_{\rm eq}(v)\,dv=
\frac{1}{\sqrt{2\pi}}\int_{-\infty}^{\infty}\!|v|\,{\rm e}^{-v^2/2}\,dv=
\left(\frac{2}{\pi}\right)^{1/2}=0.7979\:.
\label{kappa}
\end{equation}
Using $\kappa$ given by Eq.(\ref{kappa}), one can find the corresponding 
Milne's length, $l=D/\kappa=1.253/\gamma$, which has to be compared with the 
exact value, $l=-\zeta(1/2)\,D=1.46/\gamma$ (where $\zeta(z)$ is the zeta 
function \cite{Stegun}), obtained from a very sophisticated 
derivation \cite{Marsh}. The two Milne's lengths are in 
a reasonably good agreement, considering the simplicity of our derivation.   

The survival probability $S_{\rm high}(t)$ (where the subscript ''high'' 
designates the high friction regime) is :
\begin{equation}
S_{\rm high}(t)=\int_{0}^{1}\!p(x,t)\,dx\:,
\end{equation}
where $p(x,t)$ is the solution of Eq.(\ref{KKsd}) with the initial condition 
$p(x,t=0)=1$ and boundary conditions (\ref{rbca}) and (\ref{rbcb}). 
The mean lifetime, $k_{\rm high}^{-1}$, of a particle diffusing between 
partially absorbing boundaries and initially uniformly distributed on the 
interval is \cite{Bicout1}:
\begin{equation}
\frac{1}{k_{\rm high}}=\frac{1}{k_{\rm eq}}\,+\,\frac{1}{k_{\rm d}}\:,
\label{ksd}
\end{equation}
where $k_{\rm eq}$ is the rate constant in the reaction controlled limit (i.e., 
$\gamma\rightarrow 0$, $D\rightarrow \infty$ limit or $\kappa\rightarrow 0$) 
given by:
\begin{equation}
k_{\rm eq}=\kappa\,p_{\rm eq}(0)+\kappa\,p_{\rm eq}(1)=2\kappa\:.
\label{keq}
\end{equation}
The second term $k_{\rm d}$ in 
Eq.(\ref{ksd}) is the rate constant in the diffusion controlled limit 
(i.e., $\gamma\rightarrow \infty$, $D\rightarrow 0$ limit or 
$\kappa\rightarrow \infty$) which is the inverse of the mean first passage 
time of the particle to the boundaries at $x=0$ and $x=1$, averaged over 
the uniform distribution of initial positions:
\begin{equation}
k_{\rm d}^{-1}=\frac{1}{12D}=\frac{\gamma}{12}\:.
\label{kd}
\end{equation}
Plugging this and the expression in Eq.(\ref{keq}) into Eq.(\ref{ksd}), 
we find that the effective rate constant is:
\begin{equation}
\frac{1}{k_{\rm high}}=\frac{1}{2\kappa}\,+\,\frac{1}{12D}
=\frac{1}{2\kappa}\,+\,\frac{\gamma}{12}\:.
\label{khigh}
\end{equation}

\subsection{Low Friction Regime} 
\label{sec:Vdiff}

In this section we analyze the low friction regime where $\gamma$ is small 
and, hence, the particle velocity is a slowly varying variable. First, 
we reduce the Klein-Kramers equation (\ref{KK}) with the absorbing boundary 
conditions (\ref{bc}) to a diffusion equation along the velocity 
coordinate in the presence of a sink term describing the escape of the 
particle from the interval. Next, we calculate the effective rate 
constant by solving this equation. 

When $\gamma=0$, the survival probability of a particle starting at $x$ 
with initial velocity $v$ is given by Eq.(\ref{soft0}), and the lifetime 
of such a particle is given by:
\begin{equation}
\int_{0}^{\infty}\!S(t|x,v)\,dt=\frac{x}{|v|}\,{\rm H}(-v)+
\frac{(1-x)}{v}\,{\rm H}(v)\:.
\end{equation} 
Averaging this time with respect to uniform distribution of $x$ gives the 
mean lifetime, $1/K(v)$, 
\begin{equation}
\frac{1}{K(v)}=\int_{0}^{1}\!dx\int_{0}^{\infty}\!S(t|x,v)\,dt=\frac{1}{2|v|}\:.
\label{sink}
\end{equation}
$K(v)$ can be regarded as the effective rate constant for the  escape of  
a particle with initial velocity $v$. 

Let $f(v,t)$ be the probability density of the velocity. For finite $\gamma$ 
and in the absence of any reaction leading to the depletion of the 
probability density, $f(v,t)$ satisfies the differential equation,
\begin{equation}
\frac{\partial f(v,t)}{\partial t}=\gamma\,
\frac{\partial}{\partial v}\,{\rm e}^{-v^2/2}\,
\frac{\partial}{\partial v}\,{\rm e}^{v^2/2}\,f(v,t)\:.
\label{KKvd0}
\end{equation}
This equation describes the relaxation of $f(v,t)$ to the Maxwell 
distribution, $f_{\rm eq}(v)$, in Eq.(\ref{feq}). Because of the escape of 
particles from the interval, the probability density of the velocity 
decreases with time. The simplest way of taking this decrease into account 
is to incorporate a sink term into Eq.(\ref{KKvd0}) as,
\begin{equation}
\frac{\partial f(v,t)}{\partial t}=\gamma\,
\frac{\partial}{\partial v}\,{\rm e}^{-v^2/2}\,
\frac{\partial}{\partial v}\,{\rm e}^{v^2/2}\,f(v,t)\,-\,K(v)\,f(v,t)\:,
\label{KKvd}
\end{equation}
where $K(v)$ is chosen so that in the limit $\gamma \rightarrow 0$, 
the mean lifetime of the particle given in Eq.(\ref{sink}) is recovered, 
i.e., $K(v)=2|v|$.  

The survival probability $S_{\rm low}(t)$ is given by 
\begin{equation}
S_{\rm low}(t)=\int_{-\infty}^{\infty}\!f(v,t)\,dv\:,
\end{equation}
where $f(v,t)$ is the solution of Eq.(\ref{KKvd}) with the initial 
condition $f(v,t=0)=f_{\rm eq}(v)$. When $\gamma=0$, the survival 
probability obtained by solving Eq.(\ref{KKvd}) is,
\begin{equation}
S_{\rm low}(t)=\int_{-\infty}^{\infty}\!f_{\rm eq}(v)\,{\rm e}^{-K(v)\,t}\,dv
=\int_{-\infty}^{\infty}\!\frac{{\rm e}^{-v^2/2}}{\sqrt{2\pi}}
\,{\rm e}^{-2\,|v|\,t}\,dv={\rm e}^{2t^2}\,
{\rm erfc}\left(\sqrt{2}\:t\right)\:.
\label{slow}
\end{equation}
This function turns out to have exactly the same long time behavior like 
the exact survival probability given in Eq.(\ref{sinf}). As shown in Fig. 3, 
the two survival probabilities essentially coincide except for short times. 

Integration of Eq.(\ref{KKvd}) with respect to $v$ gives the time derivative 
of $S_{\rm low}(t)$ at $t=0$ as:
\begin{equation}
-\left.\frac{dS_{\rm low}}{dt}\right|_{t=0}=
\int_{-\infty}^{\infty}\!K(v)\,f_{\rm eq}(v)\,dv=2\langle|v|\rangle
=2\kappa=k_{\rm eq}\:.
\label{keqlow}
\end{equation}
One can check that $dS_{\rm low}/dt=dS_{\rm high}/dt$ at $t=0$. Another 
interesting limit of Eq.(\ref{KKvd}) is when 
$\gamma \rightarrow \infty$. In this case, the velocity distribution 
instantaneously relaxes to $f_{\rm eq}(v)$ so that the probability density 
can be written as, $f(v,t)=f_{\rm eq}(v)\,S_{\rm low}(t)$, and  the 
survival probability decays exponentially, 
$S_{\rm low}(t)={\rm e}^{-k_{\rm eq}t}$. From this it follows that the 
effective rate constant is equal to $k_{\rm eq}$ in the 
$\gamma \rightarrow \infty$ limit.

\subsubsection{Effective Rate Constant}

Let $\tau(v)$ be the lifetime of the particle initially with velocity $v$ in 
the presence of the sink term, $K(v)$. The effective rate constant in the 
low friction regime can be expressed in terms of $\tau(v)$ as
\begin{equation}
k_{\rm low}^{-1}=\int_{-\infty}^{+\infty}\!\tau(v)\,f_{\rm eq}(v)\,dv\:.
\end{equation}
The lifetime $\tau(v)$ satisfies the adjoint equation, 
\begin{equation}
\gamma\,{\rm e}^{v^2/2}\,\frac{d}{dv}\,{\rm e}^{-v^2/2}\,
\frac{d\tau(v)}{dv}\,-\,2\,|v|\,\tau(v)=-1\:.
\label{tauvd}
\end{equation}
The boundary conditions that complement this equation are: $\tau(v)$ 
vanishes as $|v|\rightarrow \infty$ and $\left.d\tau(v)/dv\right|_{v=0}=0$,  
since $\tau(v)$ is a symmetric function of $v$, i.e. $\tau(-v)=\tau(v)$.

When $\gamma$ is small, Eq.(\ref{tauvd}) can be solved approximatively 
by matching the solutions found for small and large $v$. For large velocities, 
$|v|>v^{\ast}$, where $v^{\ast}$ is a constant to be determined, 
the second term in Eq.(\ref{tauvd}) is more important than the first one 
(since $\gamma$ is small) and so $\tau(v)\simeq 1/2|v|$. At small $v$, the 
second term, proportional to $v$, can be neglected and the exponentials can 
be replaced by unity. Thus, Eq.(\ref{tauvd}) takes the form
\begin{equation}
\gamma\,\frac{d^2\tau(v)}{dv^2}=-1\:\:\:;\:\:\:
|v|\leq v^{\ast}\:.
\label{tauvd1}
\end{equation}
The solution satisfying the condition $\left. d\tau(v)/dv\right|_{v=0}=0$ is
\begin{equation} 
\tau(v)=\tau(0)-\frac{v^2}{2\gamma}\:.
\end{equation}
The time $\tau(0)$ and the velocity $v^{\ast}$ are determined from the 
condition of continuity of $\tau(v)$ and $d\tau(v)/dv$ at $v=v^{\ast}$. 
This leads to 
\begin{eqnarray}
\tau(v)=\left\{\begin{array}{lcc}
1/2v^{\ast}+(v^{\ast\,2}-v^2)/2\gamma & ; & 0\leq |v|\leq v^{\ast}\:,\\
1/2|v| & ; & |v|\geq v^{\ast}\:,
\end{array}\right.
\label{tauvd2}
\end{eqnarray}
where $v^{\ast}=(\gamma/2)^{1/3}$. 

Averaging $\tau(v)$ with respect to $f_{\rm eq}(v)$, one finds that the 
leading term of $k_{\rm low}(\gamma)$ as $\gamma\rightarrow 0$ is:
\begin{equation}
k_{\rm low} \approx 
3\sqrt{2\pi}\,\left[\ln\left(\frac{1}{\gamma}\right)\right]^{-1}
=\frac{3\pi}{2}\,
\left[\ln\left(\frac{1}{\gamma}\right)\right]^{-1}\,k_{\rm eq}\:. 
\label{klowapp}
\end{equation}
This result is {\em exact} and can be derived directly from Eq.(\ref{KK}) with 
appropriate boundary conditions. Since $k_{\rm low}=k_{\rm eq}$ as 
$\gamma\rightarrow \infty$, the following heuristic formula can be used to 
interpolate $k_{\rm low}$ over the entire range of $\gamma$:
\begin{equation}
k_{\rm low}=k_{\rm eq}\,\left\{1+\frac{2}{3\pi}\,
\ln\left[1+\frac{A}{\gamma}\right]\right\}^{-1}\:,
\label{klow}
\end{equation}
in which $A$ is the only unknown constant. It is found that 
Eq.(\ref{klow}) accurately fits exact results (obtained by solving 
Eq.(\ref{tauvd}) numerically) for $A=1.45$.

\subsection{Turnover}

So far we have derived expressions for $k_{\rm low}$ and $k_{\rm high}$ 
as functions of $\gamma$. Now, we use the VMM~\cite{Melnikov,Visscher} 
interpolation formula to obtain an analytic expression for $k$ that covers 
the entire range of $\gamma$. This leads to
\begin{equation}
\frac{k(\gamma)}{k_{\rm eq}}=\frac{k_{\rm low}(\gamma)\,k_{\rm high}(\gamma)}
{k_{\rm eq}^2}=
\left\{\left[1+\frac{2}{3\pi}\,
\ln\left(1+\frac{A}{\gamma}\right)\right]\,
\left[1+\frac{\gamma}{\sqrt{18\pi}}\right]\right\}^{-1}\:. 
\label{mmk}
\end{equation}
In order to test the theory, we have performed Langevin dynamics simulations 
to compute the exact survival probability, $S(t)$, and the effective rate 
constant, $k$, as described in Appendix~\ref{sec:Sim}. The results are 
reported in Fig.~\ref{fig2} where the closed circles represent simulation 
data, solid and dashed lines give the dependences predicted 
by Eq.(\ref{mmk}) with $A=1.45$ and $A=6$, respectively, while dot-dashed 
and long-dashed lines display $k_{\rm high}$ (Eq.(\ref{khigh})) and 
$k_{\rm low}$ (Eq.(\ref{klow})) with $A=1.45$, respectively. The dotted line 
designates the purely diffusive part of $k_{\rm high}$ (see Eq.(\ref{khigh})). 

The rate constant shows a turnover behavior as a function of $\gamma$. In the 
low friction 
regime, it goes to zero like $k\sim 1/\ln(\gamma^{-1})$, in contrast to 
the Kramers' rate constant which is proportional to $\gamma$. The theory, 
i.e. Eq.(\ref{mmk}) with $A=1.45$, describes simulation results reasonably  
well. However, Eq.(\ref{mmk}) with $A=6$ is in excellent agreement with 
the simulation data as shown by the dashed line through the data in Fig. 2. 
Equation (\ref{mmk}) with $A=6$ can  be regarded as exact for practical 
purposes. The rate constant $k$ increases as $\gamma$ gets larger, attains 
its maximum value (smaller than $k_{\rm eq}$) close to $\gamma\simeq 1$, 
turns over and then decreases like $k\sim \gamma^{-1}$ as 
$\gamma\rightarrow \infty$. In this friction regime, the theory is in 
excellent agreement with simulation results. Comparison between the 
dot-dashed and dotted lines (i.e., diffusion with perfectly absorbing 
boundary conditions) shows how results are improved by using radiation 
boundary conditions.  

In Fig. 3, $S(t)$ is plotted versus $t$ for various values of $\gamma$. 
When $\gamma=0$, the survival probability is strongly non-exponential and 
goes as $t^{-1}$ at long times (see Eq.(\ref{sinf})). For comparison we have 
also plotted $S_{\rm low}(t)$ given in Eq.(\ref{slow}). As one can see, 
$S_{\rm low}(t)$ is slightly below $S(t)$ at short times (since the slopes 
at $t=0$ are $2\kappa$ and $\kappa$, respectively), but they become 
indistinguishable from each other at longer times. 

For $\gamma=\infty$, $S(t)=1$ since the particle cannot reach the absorbing 
boundaries. When $\gamma=1$, the decay of $S(t)$ is close to exponential but 
with a slope smaller than $k_{\rm eq}=2\kappa$ represented by 
${\rm e}^{-2\kappa t}$ (dot-dashed line). As $\gamma$ becomes smaller than 
unity, $S(t)$ becomes more and more non-exponential and decays very slowly. 
When $\gamma$ becomes greater than unity, $S(t)$ is almost an exponential 
with the slope $k_{\rm high}$ given in Eq.(\ref{khigh}). 

To summarize, the main result of this section is the expression given in 
Eq.(\ref{mmk}) for the effective rate constant for a particle moving between 
two absorbing boundaries. It describes the turnover behavior of the rate 
constant as a function of $\gamma$. We will use this result in the next 
section to obtain the turnover of the steady-state rate constant in the 
trapping problem.

\section{Steady-State Rate Constant for the Trapping Problem in One Dimension}
\label{sec:SSR}
\setcounter{equation}{0}

Consider an ensemble of traps (absorbing points) uniformly distributed on an 
infinite line. If $c$ is the concentration of traps (i.e., the number of traps 
per unit of length) and $L$ denotes the distance between neighboring traps, 
the distribution of $L$ is given by, $c\,{\rm e}^{-cL}$. A Langevin particle 
is injected at any point of the line with equal probability. The probability 
density that a particle is injected on the interval of length $L$ is 
$c^2L\,{\rm e}^{-cL}$. 
Let $s(t|L)$ be the survival probability of the particle moving within the 
interval $L$ averaged over uniform distribution of the initial position within 
the interval and over the equilibrium distribution of the initial velocity. 
Averaging $s(t|L)$ with respect to $L$, gives the survival probability 
of the particle in the presence of static traps at concentration $c$:
\begin{equation}
{\cal S}(t)=c^2\int_{0}^{\infty}\!L\,{\rm e}^{-cL}\,
s(t|L)\,dL\:.
\label{soft}
\end{equation}
Before analyzing the turnover of the steady-state rate constant as a function 
of $\gamma$, we briefly consider the asymptotic long time behavior of 
${\cal S}(t)$, which is determined by particles located in large intervals 
free from traps. Such a particle changes its 
velocity many times before being trapped. As a consequence, the particle 
motion is essentially diffusive and its averaged survival probability 
approaches zero as ${\rm e}^{-t^{1/3}}$ at very long times \cite{Bala}. 
 
The steady-state rate constant, $k_{\rm ss}$, is expressed in terms of the 
survival probability as, 
\begin{equation}
k_{\rm ss}^{-1}=c\int_{0}^{\infty}\!{\cal S}(t)\,dt
=c^3\int_{0}^{\infty}\!dt\int_{0}^{\infty}\!L\,{\rm e}^{-cL}\,
s(t|L)\,dL\:.
\end{equation}
Changing variables so as to express $s(t|L)$ in terms of the survival 
probability on a unit interval and carrying out the integration over the time, 
we have:
\begin{eqnarray}
k_{\rm ss}(\varepsilon)=
\left[\int_{0}^{\infty}\!\frac{z^2\,{\rm e}^{-z}}
{k(\varepsilon z)}\,dz\right]^{-1}\:\:\:;\:\:\:
\varepsilon=\frac{\gamma}{c}\:,
\end{eqnarray}
where $k(\varepsilon z)$ is the effective rate constant given in 
Eq.(\ref{mmk}). Using this expression we find:
\begin{equation}
\frac{k_{\rm ss}(\varepsilon)}{\kappa}=
\left[1+\frac{\varepsilon}{\sqrt{2\pi}}+
\frac{1}{3\pi}\,h_1\left(\frac{A}{\varepsilon}\right)+
\frac{\varepsilon}{9\pi\sqrt{2\pi}}
\,h_2\left(\frac{A}{\varepsilon}\right)\right]^{-1}\:,
\label{kss}
\end{equation}
in which the functions $h_1(y)$ and $h_2(y)$ are defined as,
\setcounter{xeq}{\value{equation}}
\addtocounter{xeq}{1}
\renewcommand{\theequation}{\arabic{section}.\arabic{xeq}\alph{equation}}
\setcounter{equation}{0}
\begin{eqnarray}
h_1(y) & = &  2\gamma_{\rm E}+2\ln(y)-y+\left[y^2-2y+2\right]\,{\rm e}^{y}
\,{\rm E}_1(y)\:, \\
h_2(y) & = & 6\gamma_{\rm E}+6\ln(y)-4y+y^2-\left[y^3-3y^2+6y-6\right]\,
{\rm e}^{y}\,{\rm E}_1(y)\:,
\end{eqnarray}
where $\gamma_{\rm E}=0.577216$ is the Euler's constant and,  
${\rm E}_1(y)=\int_{y}^{\infty}\!z^{-1}\,{\rm e}^{-z}\,dz$, is the exponential 
integral function \cite{Stegun}. 
\renewcommand{\theequation}{\arabic{section}.\arabic{equation}}
\setcounter{equation}{\value{xeq}}

Equation (\ref{kss}) predicts a turnover behavior of the steady-state rate 
constant as a function of $\varepsilon$. When $\varepsilon\rightarrow 0$, 
$k_{\rm ss}$ approaches zero as
\begin{equation}
\frac{k_{\rm ss}(\varepsilon)}{\kappa}\simeq 
\frac{3\pi}{2}\,\left[\ln\left(\frac{B}{\varepsilon}\right)\right]^{-1}
\:\:\:;\:\:\:B=A\,\exp\left\{\frac{3\pi-3+2\gamma_{\rm E}}{2}\right\}\:.
\label{kss0}
\end{equation}
In the opposite limit when $\varepsilon\rightarrow \infty$, the 
steady-state rate constant tends to zero like, 
\begin{equation}
\frac{k_{\rm ss}(\varepsilon)}{\kappa}
\simeq \frac{\sqrt{2\pi}}{\varepsilon}\:.
\label{kssinf}
\end{equation}
In Fig. 4 the $\varepsilon$-dependence of $k_{\rm ss}(\varepsilon)$ given in 
Eq.(\ref{kss}) is shown for $A=6$ (solid curve) and $A=1.45$ (dashed curve). 
The former may be considered as exact since it is obtained with $A=6$, which 
provides an almost perfect fit to simulation data for the unit interval 
(see Fig. 2). The dashed curve shows the dependence obtained on the basis of 
the theory developed in Section~\ref{sec:Vdiff}.  Although both curves have 
the same qualitative behavior there is clearly a room for improvement. In 
what follows we develop another approach 
which allows one to calculate $k_{\rm ss}(\varepsilon)$ in a space of 
arbitrary dimension. We will use the dependence shown by the solid curve in 
Fig. 4 in order to test the approach in one dimension. It will be shown 
that the new theory predicts the dependence of $k_{\rm ss}(\varepsilon)$ 
which is in better agreement with the exact result (solid curve in Fig. 4) 
than does the theory developed in Section~\ref{sec:Vdiff} (dashed curve 
in Fig. 4). 

\section{New Approach}
\label{sec:newapp}
\setcounter{equation}{0}

We now introduce a new and simpler approach to calculate the steady-state 
rate constant for the trapping problem which  has the great advantage of 
being generalizable to any dimensions. Moreover, this approximate approach 
is expected to improve as the dimension of the space increases. 
As we will see below, even in one dimension, the new approach is already 
in better agreement with the simulations than does the previous one 
(compare Figs. 4 and 5).

The basic idea of this new approach is to apply the VMM interpolation 
formula directly to the steady-state rate constants found in the low 
and high friction regimes. In the low friction regime the rate constant 
will be obtained by using a many-particle generalization of the approach 
presented in Section~\ref{sec:Vdiff}. In the high friction regime we will 
assume that the steady-state rate constant for the trapping problem is 
equal to the rate constant for the target problem. This approximation is 
quite good in one dimension and is virtually exact in three dimensions.

\subsection{Target Problem}
\label{sec:target}

The survival probability of the target particle (i.e., the $A$ in 
the reaction $A+B\,\longrightarrow\,B$ with $D_A=0$), is given by
\begin{equation}
{\cal S}_{\rm targ}(t)=\exp\left\{-c\int_{0}^{t}k_{\rm targ}(t')\,dt'\right\}\:,
\end{equation}
where $c$ is the concentration of mobile traps (i.e., the $B$'s) and 
$k_{\rm targ}(t)$ is the time-dependent rate coefficient. The steady-state 
rate constant is \cite{Szabo}
\begin{equation}
k_{\rm targ}^{-1}=c\int_{0}^{\infty}\!{\cal S}_{\rm targ}(t)\,dt\:.
\end{equation}
The exact time-dependent rate coefficient $k_{\rm targ}(t)$ is obtained by 
solving the Klein-Kramers equation with appropriate boundary conditions. 
This is an extremely complicated problem \cite{Harris} and here we will 
approximate $k_{\rm targ}(t)$ by the Collins-Kimball rate constant, 
$k_{\rm CK}(t)$, which is obtained by solving the diffusion equation with 
the radiation boundary condition at contact, in which the intrinsic rate 
constant is chosen so that $k_{\rm CK}(0)$ is exact.

First, we note that $k_{\rm targ}(t=0)$ can be found exactly for any friction. 
The product $ck_{\rm targ}(0)$ is just the collision frequency, therefore
\begin{eqnarray}
k_{\rm targ}(0)=\kappa=\left\{\begin{array}{ccc}
\langle|v|\rangle=\sqrt{2/\pi} & ; & d=1\:, \\
\pi R^2\,\langle|v|\rangle=\sqrt{8\pi}\:R^2 & ; & d=3\:,
\end{array}\right.
\label{ktarg0}
\end{eqnarray}
where $R$ is the contact radius. In addition, when the friction vanishes, 
$\gamma=0$, $k_{\rm targ}(t)=k_{\rm targ}(0)$ and thus the rate constant 
is given by Eq.(\ref{ktarg0}) for all times. As the friction tends to 
infinity, the rate constant is given by the Smoluchowski theory. Now, 
we consider the Collins-Kimball rate coefficient, $k_{\rm CK}(t)$, 
obtained for the diffusion constant $D=1/\gamma$ and the intrinsic rate 
constant $\kappa$ appearing in the boundary conditions equal to 
$k_{\rm targ}(0)$ given  in Eq.(\ref{ktarg0}). This Collins-Kimball 
rate coefficient has the following properties: 
{\em i)} $k_{\rm CK}(0)=k_{\rm targ}(0)$ at any friction, {\em ii)} 
when the friction vanishes,  we have $k_{\rm CK}(t)=k_{\rm targ}(0)$ 
for all times, and {\em iii)} at high friction $k_{\rm CK}(t)$ coincides 
with the Smoluchowski rate constant except for very short times where 
it is equal to $k_{\rm targ}(0)$. In the view of these properties 
we approximate $k_{\rm targ}(t)$ by $k_{\rm CK}(t)$. As a consequence, we have 
${\cal S}_{\rm targ}(t)={\cal S}_{\rm CK}(t)$ and 
\begin{equation}
k_{\rm targ}^{-1}=c\int_{0}^{\infty}\!{\cal S}_{\rm CK}(t)\,dt\:.
\label{ktargck}
\end{equation}

\subsection{Trapping Problem in the Low Friction Regime}
\label{sec:trap}

When the friction vanishes, the particle moves with a constant velocity. Its 
survival probability is equal to the probability that the region visited by 
the particle in time $t$ is free from traps. Since the traps are uniformly 
distributed in space, the probability that a region is free from traps is 
given by the Poisson distribution. The survival probability of a particle 
moving with a velocity $v$ is  
\begin{eqnarray}
{\cal S}(t|v)=\left\{\begin{array}{lcc}
\exp(-c|v|t) & ; & d=1\:, \\
\exp(-c\pi R^2|v|t)) & ; & d=3\:.
\end{array}\right.
\label{stv}
\end{eqnarray}
Assuming equilibrium distribution of the initial velocity, the 
probability density of the velocity at time $t$ is 
\begin{equation}
f(v,t)={\cal S}(t|v)\,f_{\rm eq}(v)\:.
\label{fvt}
\end{equation}
Integrating this with respect to velocity, we find the survival probability 
in the ballistic regime
\begin{equation}
{\cal S}_{\rm ball}(t)=\int\!f(v,t)\,dv=\int\!{\cal S}(t|v)\,f_{\rm eq}(v)\,dv\:.
\end{equation}
This result is exact. In one dimension it leads to
\begin{equation}
{\cal S}_{\rm ball}(t)=\left(\frac{2}{\pi}\right)^{1/2}
\int_{0}^{\infty}\!\exp\left\{-cvt-\frac{v^2}{2}\right\}\,dv
=\exp\left(\frac{c^2t^2}{2}\right)\,
{\rm erfc}\left(\frac{ct}{\sqrt{2}}\right)\:.
\end{equation}
This expression can also be obtained using $S(t)$ in Eq.(\ref{sexact}) 
in the relation for ${\cal S}(t)$ given in Eq.(\ref{soft}). In three 
dimensions ${\cal S}_{\rm ball}(t)$ is
\begin{eqnarray}
{\cal S}_{\rm ball}(t) & = & \left(\frac{2}{\pi}\right)^{1/2}
\int_{0}^{\infty}\!v^2\,\exp\left\{-c\pi R^2vt-\frac{v^2}{2}\right\}\,dv 
\nonumber \\
& = & \left(1+\frac{\pi \tau^2}{8}\right)\,\exp\left(\frac{\pi \tau^2}{16}
\right)\,{\rm erfc}\left(\frac{\sqrt{\pi}}{4}\,\tau\right)-\frac{\tau}{2}
\:\:\:;\:\:\:\tau=\kappa ct=\sqrt{8\pi}\:cR^2t\:,
\end{eqnarray}
where  we have used the relation in Eq.(\ref{ktarg0}) for $\kappa$ in three 
dimensions, i.e.  $\kappa=\sqrt{8\pi}\:R^2$.

Equations (\ref{stv}) and (\ref{fvt}) show that particles with large 
velocities react more rapidly and therefore the higher the velocity the 
faster the probability density decays. When the friction is non-zero, due 
to the interaction with environment, the probability density, depleted by 
the reaction, relaxes to the Maxwell distribution. At small friction 
$f(v,t)$ satisfies 
\begin{equation}
\frac{\partial f(v,t)}{\partial t}=\left[{\cal L}(v)-k(v)\right]\,f(v,t)\:,
\label{dfvt}
\end{equation}
where ${\cal L}(v)$ describes the relaxation of the velocity distribution 
in $d$-dimensions 
\begin{equation}
{\cal L}(v)=\frac{\gamma}{v^{d-1}}\,\frac{\partial}{\partial v}
v^{d-1}\,{\rm e}^{-v^2/2}\,\frac{\partial}{\partial v}\,{\rm e}^{v^2/2}\:,
\end{equation}
and the sink term $k(v)$ describes the reaction. In one and three 
dimensions the sink term is 
\begin{eqnarray}
k(v)=\left\{\begin{array}{lcc}
c|v| & ; & d=1\:, \\
\pi R^2\,c\,|v| & ; & d=3\:.
\end{array}\right.
\end{eqnarray}
When $\gamma=0$, ${\cal L}(v)$ vanishes, and $f(v,t)$, obtained from 
Eq.(\ref{dfvt}) with initial condition $f(v,0)=f_{\rm eq}(v)$, coincides 
with Eq.(\ref{fvt}).

The steady-state rate constant can be obtained by averaging the mean lifetime 
$\tau(v)$ of the particle with the initial velocity $v$, with respect to the 
equilibrium distribution of the initial velocity:
\begin{equation}
k_{\rm low}^{-1}=\int_{0}^{\infty}\!v^{d-1}\,f_{\rm eq}(v)\tau(v)\,dv\:\:\:;
\:\:\:\int_{0}^{\infty}\!v^{d-1}\,f_{\rm eq}(v)\,dv=1\:.
\label{ktauv}
\end{equation}
The lifetime $\tau(v)$ can be determined by solving the backward equation 
\begin{equation}
\left[{\cal L}^{\dag}(v)-k(v)\right]\,\tau(v)=-1\:,
\label{dtauv}
\end{equation}
where ${\cal L}^{\dag}(v)$ is the adjoint operator
\begin{equation}
{\cal L}^{\dag}(v)=\frac{\gamma}{v^{d-1}}\,{\rm e}^{v^2/2}\,
\frac{d}{dv}v^{d-1}\,{\rm e}^{-v^2/2}\,\frac{d}{dv}\:.
\end{equation}

In summary, to find the steady-state rate constant for the trapping problem 
one has to calculate the rate constants $k_{\rm low}$ and $k_{\rm high}$ 
corresponding to low and high friction regimes and then use the VMM interpolation 
formula. The rate constant $k_{\rm low}$ is given by Eq.(\ref{ktauv}) where 
$\tau(v)$ is obtained by solving Eq.(\ref{dtauv}). The rate $k_{\rm high}$ is 
assumed to be approximatively equal to the steady-state rate constant for the 
target problem $k_{\rm targ}$ given in Eq.(\ref{ktargck}). In the following 
section we use this approach to calculate the steady-state rate constant 
over the entire range of friction in one and three dimensions.

\section{Steady-State Rate Constant}
\label{sec:SS}
\setcounter{equation}{0}

\subsection{One-Dimensional Case}
\label{sec:1d}

We begin with the calculation of $k_{\rm targ}$ based on Eq.(\ref{ktargck}) 
using the survival probability
\begin{equation}
{\cal S}_{\rm CK}(t)=\exp\left\{-c\left[\sqrt{\frac{16Dt}{\pi}}+
\frac{4D}{\kappa}\,\left(\exp\left(\frac{\kappa^2t}{4D}\right)\,
{\rm erfc}\left(\sqrt{\frac{\kappa^2t}{4D}}\right)-1\right)
\right]\right\}\:,
\end{equation}
where $D=1/\gamma$ and $\kappa$ is given in Eq.(\ref{ktarg0}). As a result, we 
obtain
\begin{equation}
\frac{k_{\rm targ}(\varepsilon)}{\kappa}=\left\{\frac{\sqrt{8\pi}}{\varepsilon}
\int_{0}^{\infty}\!\exp\left\{-\frac{\sqrt{8\pi}}{\varepsilon}\left[
\sqrt{\frac{4z}{\pi}}+{\rm e}^z\,{\rm erfc}\left(\sqrt{z}\right)-1\right]\right\}
\,dz\right\}^{-1}\:\:\:;\:\:\:\:\varepsilon=\frac{\gamma}{c}\:.
\label{ksshigh}
\end{equation}
The ratio $k_{\rm targ}(\varepsilon)/\kappa$ as a function of $\varepsilon$ 
is represented by the dash-dotted line in Fig. 5. One can see that 
$k_{\rm targ}(\varepsilon)/\kappa$ monotonically decreases with $\varepsilon$ 
from unity at $\varepsilon=0$ to zero as $\varepsilon\rightarrow\infty$. For 
sufficiently large $\varepsilon$ the steady-state rate constant approaches the 
one predicted by the Smoluchowski theory and tends to zero like $1/\varepsilon$.

To calculate $k_{\rm low}$ by Eq.(\ref{ktauv}) we have to determine $\tau(v)$ 
solving Eq.(\ref{dtauv}) which in one dimension is
\begin{equation}
\gamma\,{\rm e}^{v^2/2}\,
\frac{d}{dv}\,{\rm e}^{-v^2/2}\,
\frac{d\tau(v)}{dv}-c|v|\tau(v)=-1\:.
\end{equation}
By changing variables this equation can be reduced to Eq.(\ref{tauvd}). 
This allows us to use the solution in Eq.(\ref{klow}) and to eventually 
obtain 
\begin{equation}
\frac{k_{\rm low}(\varepsilon)}{\kappa}=\left[1+\frac{2}{3\pi}\,
\ln\left(1+\frac{A}{2\varepsilon}\right)\right]^{-1}\:,
\label{ksslow}
\end{equation}
where $A=1.45$. The dashed line in Fig. 5 represents 
$k_{\rm low}(\varepsilon)/\kappa$ as a function of $\varepsilon$. The ratio 
monotonically increases with $\varepsilon$ from zero in the ballistic regime 
($\varepsilon=0$) to unity as $\varepsilon\rightarrow\infty$. The rate 
constant $k_{\rm low}(\varepsilon)$ vanishes as $1/\ln(1/\varepsilon)$ as 
$\varepsilon\rightarrow 0$. 

We now use the VMM interpolation formula to get the steady-state rate constant 
for the trapping problem for the entire range of friction (i.e. of the parameter 
$\varepsilon=\gamma/c$)
\begin{equation}
\frac{k_{\rm trap}(\varepsilon)}{\kappa}=
\frac{k_{\rm low}(\varepsilon)\,k_{\rm high}(\varepsilon)}
{\kappa^2}=\frac{k_{\rm low}(\varepsilon)\,k_{\rm targ}(\varepsilon)}
{\kappa^2}\:.
\label{mmktrap}
\end{equation}
Using the expressions in Eqs.(\ref{ksshigh}) and (\ref{ksslow}) we find
\begin{eqnarray}
& & \frac{k_{\rm trap}(\varepsilon)}{\kappa}= \nonumber \\
& & \left\{\left[1+\frac{2}{3\pi}\,
\ln\left(1+\frac{1.45}{2\varepsilon}\right)\right]\,
\frac{\sqrt{8\pi}}{\varepsilon}
\int_{0}^{\infty}\!\exp\left\{-\frac{\sqrt{8\pi}}{\varepsilon}\left[
\sqrt{\frac{4z}{\pi}}+{\rm e}^z\,{\rm erfc}\left(\sqrt{z}\right)-1\right]\right\}
\,dz\right\}^{-1}\:.
\label{ktrap}
\end{eqnarray}
This function is represented by the long dashed line in Fig. 5. It should be 
compared with the exact solution found in Section~\ref{sec:SSR} which is 
represented by the solid line in Fig. 5. The approximate result in 
Eq.(\ref{ktrap}) is overall closer to the exact dependence than the solution 
found in Section~\ref{sec:SSR} (see Fig. 4 and related discussion). 
However, Eq.(\ref{ktrap}) is no longer exact in the diffusive limit 
($\varepsilon\rightarrow \infty$). This is because the steady state rate 
constant of the trapping and target problems differ \cite{Szabo}. However, 
this difference is negligible in three dimensions.

\subsection{Three-Dimensional Case}
\label{sec:3d}

Like in the  one-dimensional case, we use the Collins-Kimball survival 
probability in Eq.(\ref{ktargck}) in order to calculate $k_{\rm targ}$. 
In three dimensions
\begin{equation}
{\cal S}_{\rm CK}(\tau)=\exp\left\{-\frac{1}{1+\varepsilon}\,\left[\tau+
\frac{\varepsilon}{\alpha}\,\left(\sqrt{\frac{4\alpha\tau}{\pi}}+
{\rm e}^{\alpha\tau}\,{\rm erfc}\left(\sqrt{\alpha\tau}\right)-1\right)
\right]\right\}\:\:\:;\:\:\:\tau=c\kappa t\:,
\label{stck}
\end{equation}
where $\kappa$ is given by Eq.(\ref{ktarg0}) and 
\begin{equation}
\varepsilon=\frac{\gamma\kappa}{4\pi R}=\frac{\gamma R}{\sqrt{2\pi}}\:\:\:;\:\:\:
\alpha=\frac{(1+\varepsilon)^2}{3\varepsilon\phi}\:,
\end{equation}
where $\phi=4\pi R^3c/3$ is the volume fraction of traps. Substituting  
${\cal S}_{\rm CK}(\tau)$ into Eq.(\ref{ktargck}) we obtain
\begin{equation}
\frac{k_{\rm targ}(\varepsilon,\phi)}{\kappa}=
\left[\int_{0}^{\infty}\!{\cal S}_{\rm CK}(\tau)\,d\tau\right]^{-1}\:.
\label{ktargck1}
\end{equation}
In three dimensions the steady-state rate constant is a function of two 
variables $\varepsilon$ and $\phi$. For small As $\phi$, Eq.(\ref{ktargck1}) 
reduces approximatively to 
\begin{equation}
\frac{k_{\rm targ}(\varepsilon,\phi)}{\kappa}\simeq \frac{1}{1+\varepsilon}\,
\left[1+\sqrt{\frac{3\phi \varepsilon^3}{(1+\varepsilon)^3}}\right]\:.
\label{ktargck2}
\end{equation}

Next we find the mean lifetime $\tau(v)$ by solving Eq.(\ref{dtauv}) which in 
three dimensions has the form
\begin{equation}
\frac{\gamma}{v^2}\,{\rm e}^{v^2/2}\,
\frac{d}{dv}\,\left[v^2\,{\rm e}^{-v^2/2}\,
\frac{d\tau(v)}{dv}\right]-\pi R^2cv\tau(v)=-1\:.
\label{dtauv3}
\end{equation}
Introducing an effective diffusion constant 
$\lambda=\gamma/c\kappa=2\pi\varepsilon/3\phi$, we can rewrite 
Eq.(\ref{dtauv3}) as 
\begin{equation}
\frac{\lambda}{v^2}\,{\rm e}^{v^2/2}\,
\frac{d}{dv}\,\left[v^2\,{\rm e}^{-v^2/2}\,
\frac{d\tau(v)}{dv}\right]-\sqrt{\frac{\pi}{8}}\:v\tau(v)=-\frac{1}{c\kappa}\:.
\label{dtauv4}
\end{equation}
This equation can be solved for small and large $\lambda$. When 
$\lambda\rightarrow\infty$ (instantaneous relaxation of the velocity 
distribution) $\tau(v)=1/c\kappa$ for all $v$. In the opposite limiting 
case of small $\lambda$ the first term in the right-hand side of 
Eq.(\ref{dtauv4}) can be neglected except for very small values of 
$v$. However, when $v$ is small the exponentials in the first term in 
Eq.(\ref{dtauv4}) can be replaced by unity. Thus at small $\lambda$, 
$\tau(v)$ satisfies 
\begin{equation}
\frac{\lambda}{v^2}\,\frac{d}{dv}\,\left[v^2\,\frac{d\tau(v)}{dv}\right]
-\sqrt{\frac{\pi}{8}}\:v\tau(v)=-\frac{1}{c\kappa}\:.
\label{dtauv5}
\end{equation}
A solution of this equation that is finite at $v=0$ and vanishes as 
$v\rightarrow\infty$ is
\begin{equation}
\tau(v)=\sqrt{\frac{8}{\pi}}\,\left[1-
\frac{{\rm Ai}\left(\pi^{1/6}v/2^{1/2}\lambda^{1/3}\right)}
{{\rm Ai}(0)}\right]\,\frac{1}{c\kappa v}\:,
\label{taunv}
\end{equation}
where ${\rm Ai}(z)$ is the Airy function \cite{Stegun}. 

Once $\tau(v)$ is known one can find $k_{\rm low}$ by Eq.(\ref{ktauv}). 
Using $\tau(v)$ in Eq.(\ref{taunv}) we find that as 
$\lambda\rightarrow 0$, $k_{\rm low}$ is given by 
\begin{equation}
\frac{k_{\rm low}(\lambda)}{\kappa}\simeq \frac{\pi}{4}+
\frac{3^{5/6}}{4\pi^{1/3}}\,\left[\Gamma(2/3)\right]^2\,\lambda^{2/3}\:.
\label{kll}
\end{equation}
This expression shows that in three dimensions $k_{\rm low}$ remains finite 
in the ballistic regime ($\lambda=0$), in contrast to the one-dimensional 
case where $k_{\rm low}$ vanishes as friction goes to zero. Since 
$\lambda=2\pi\varepsilon/3\phi=\gamma/c\kappa$, Eq.(\ref{kll}) shows that 
the rate constant increases as $\gamma^{2/3}$ for small $\gamma$.

As $\lambda\rightarrow\infty$, $\tau(v)=1/c\kappa$ for all $v$ and the ratio 
$k_{\rm low}(\varepsilon)/\kappa$ becomes unity. To interpolate between 
the two limiting cases of small and large $\lambda$ we use the formula
\begin{equation}
\frac{k_{\rm low}(\lambda)}{\kappa}=\frac{\frac{\pi}{4}+a\lambda^{2/3}}
{1+a\lambda^{2/3}}\:\:\:;\:\:\:
a=\frac{3^{5/6}\left[\Gamma(2/3)\right]^2}{\pi^{1/3}(4-\pi)}=3.643\:.
\label{ksslow3}
\end{equation}
To test this formula, we calculate $k_{\rm low}/\kappa$ by numerically 
solving Eq.(\ref{dtauv4}) and then carrying out the integration with 
respect to $v$ in Eq.(\ref{ktauv}). The result is reported in Fig. 6 which 
shows good agreement between the dependence predicted by Eq.(\ref{ksslow3}) 
(solid line) and the numerical data (closed circles). The dashed line 
shows the dependence predicted by Eq.(\ref{kll}) for small $\varepsilon$.

To cover the entire range of friction we again use the VMM interpolation 
formula
\begin{equation}
\frac{k_{\rm trap}(\varepsilon,\phi)}{\kappa}=
\frac{k_{\rm low}(\lambda)\,k_{\rm targ}(\varepsilon,\phi)}
{\kappa^2}\:,
\label{mmktrap3}
\end{equation}
where $k_{\rm low}(\lambda)/\kappa$ and 
$k_{\rm targ}(\varepsilon,\phi)/\kappa$ are 
given in Eqs.(\ref{ksslow3}) and (\ref{ktargck1}), respectively. When 
$\phi\ll 1$, the second term in the right hand side of Eq.(\ref{mmktrap3}) 
becomes independent of $\phi$ and is simply given by the first term in 
Eq.(\ref{ktargck2}). Thus, in the $\phi\ll 1$ limit, Eq.(\ref{mmktrap3}) 
reduces to
\begin{equation}
\frac{k_{\rm trap}(\varepsilon,\phi)}{\kappa}=
\frac{\frac{\pi}{4}+b\left(\varepsilon/\phi\right)^{2/3}}
{(1+\varepsilon)\,\left[1+b\left(\varepsilon/\phi\right)^{2/3}\right]}\:,
\end{equation}
where 
\begin{equation}
b=\left(\frac{2\pi}{3}\right)^{2/3}a=
\frac{3^{1/6}(4\pi)^{1/3}\left[\Gamma(2/3)\right]^2}{(4-\pi)}=5.964\:.
\end{equation}

Figure 7 displays the dependence of $k_{\rm trap}(\varepsilon,\phi)/\kappa$ 
(solid lines) as a function of $\varepsilon$ for two values of $\phi$. The 
dashed and dot-dashed lines represent $k_{\rm low}/\kappa$ and 
$k_{\rm targ}/\kappa$, respectively, for the same $\phi$. The rate constant 
exhibits a turnover behavior as a function of $\varepsilon$, however, the 
turnover in three dimensions depends on the volume fraction and is much 
less pronounced than in one dimension.

\section{Summary}

In this paper we have generalized the standard theory of diffusion 
controlled reactions to the case when the reactants diffuse in both 
coordinate and velocity space (i.e., they undergo Langevin rather than 
Brownian dynamics). We have developed an approximate theory of the 
steady state rate constant for both the target and trapping problems. 
This theory was tested against accurate results obtained from simulations 
for the trapping problem in one dimension and is expected to work even 
better in higher dimensions.

The key finding was that the steady state rate constant for the trapping 
(but not for the target) problem exhibits a turnover behavior as a 
function of the friction coefficient for Langevin dynamics. For Brownian 
particles the rate constants for both problems decrease monotonically as 
the friction increases. The physical explanation of this turnover is that 
for the trapping problem in the ballistic regime, a particle with near 
zero velocity can survive for a very long time. In one dimension, where 
the most probable velocity is zero, the mean lifetime is actually 
infinite and thus the rate constant is zero in this limit. Increasing 
the friction increases the rate because particles with initial velocities 
close to zero can be speeded up by random forces. In three dimensions, 
where the most probable velocity is finite, the rate constant is also 
finite in the ballistic limit and hence the turnover is less pronounced. 
Since we have assumed that the particles react on first contact, there 
is no energy barrier to reaction. Thus the friction dependence of the 
trapping steady state rate constant represents a simple physical 
example of turnover in activationless rate processes.

\newpage 

\appendix
\renewcommand{\theequation}{\Alph{section}.\arabic{equation}}

\section{The Simulation Procedure}
\label{sec:Sim}
\setcounter{equation}{0}

Simulations were performed using the discretized version of Eq.(\ref{lang})  
\begin{equation}
x_{n+1}=x_n+{\rm e}^{-\gamma\Delta}\,(x_n-x_{n-1})+X_n\:,
\label{langsim}
\end{equation}
where $\Delta$ is the time step and the Gaussian random noise $X_n$ 
(related to $R_n$ by, $X_n=[\Delta (1-{\rm e}^{-\gamma\Delta})/\gamma]\,R_n$) is 
defined by the moments,
\begin{equation}
\langle X_n\rangle=0\:\:\:\mbox{and}\:\:\:
\langle X_nX_{n'}\rangle
=\left(\frac{2\Delta}{\gamma}\right)\,(1-{\rm e}^{-\gamma\Delta})^2\,
\delta_{nn'}\:.
\end{equation}
For each trajectory, the initial position $x_0$ is generated from the uniform 
distribution between $0$ and $1$ and the initial velocity $v_0$ from the 
centered Gaussian distribution of unit standard deviation. From this, the 
first position $x_1$ at the first step is then calculated as:
\begin{equation}
x_1=x_0+v_0\,\Delta\:,
\end{equation}
and the next positions $x_n$ are generated according to the algorithm in 
Eq.(\ref{langsim}). 
To simulate the absorbing boundary conditions, each trajectory starting at 
$x_0$ ($0<x_0<1$) at time $t=0$ is terminated at time $t_i=n \Delta$ when 
either the condition $x_n\leq 0$ or $x_n\geq 1$ is satisfied for the 
first time. The first passage time $t_i$ and the survival probability $S_i(t)$ 
(defined as $S_i(t)=1$ for all $t<t_i$ and $S_i(t)=0$ otherwise) for this given 
trajectory are recorded. The survival probability, $S(t)$, and the 
effective rate constant, $k$, (i.e., the inverse of the mean lifetime) are then 
obtained by averaging over a large number of trajectories:
\begin{equation}
S(t)=\frac{1}{N}\sum_{i=1}^{N}S_i(t)\:\:\:\:\mbox{and}\:\:\:\:
k^{-1}=\frac{1}{N}\sum_{i=1}^{N}t_i\:.
\end{equation}
For all simulations reported in this paper we used the time step 
$\Delta=10^{-5}$ for $10^{-4}\leq \gamma\leq 10^3$, and $N=10^5$ 
trajectories were used to perform the averages.

\newpage


\begin{figure}[ht]
\centerline{\psfig{figure=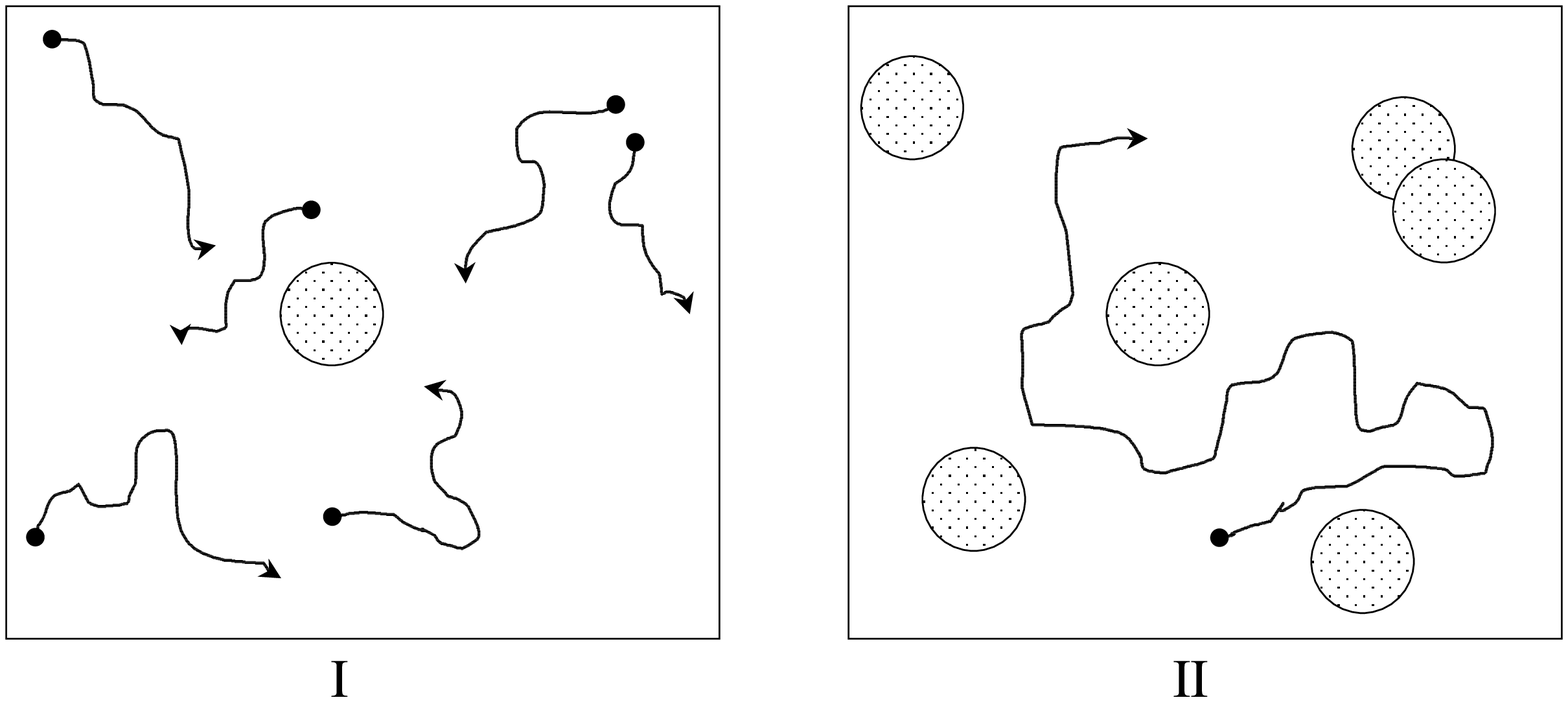,width=4.0in,angle=0}}
\vspace{-0.75cm}
\caption{Sketch of the reaction $A+B\,\longrightarrow\,B$ between a 
single $A$ with many $B$'s where the circles show the contact radius for the 
reaction. In the target problem (panel I) a single $A$ is fixed in space 
and surrounded by mobile $B$'s while in the trapping problem (panel II) the 
$B$'s are immobile and the $A$ is mobile. In this latter case the static 
traps $B$'s are distributed according to the Poisson distribution and 
overlaps may occur.}
\label{fig1}
\end{figure}

\vspace{0.25cm}

\begin{figure}[ht]
\vspace{0.3cm}
\centerline{\psfig{figure=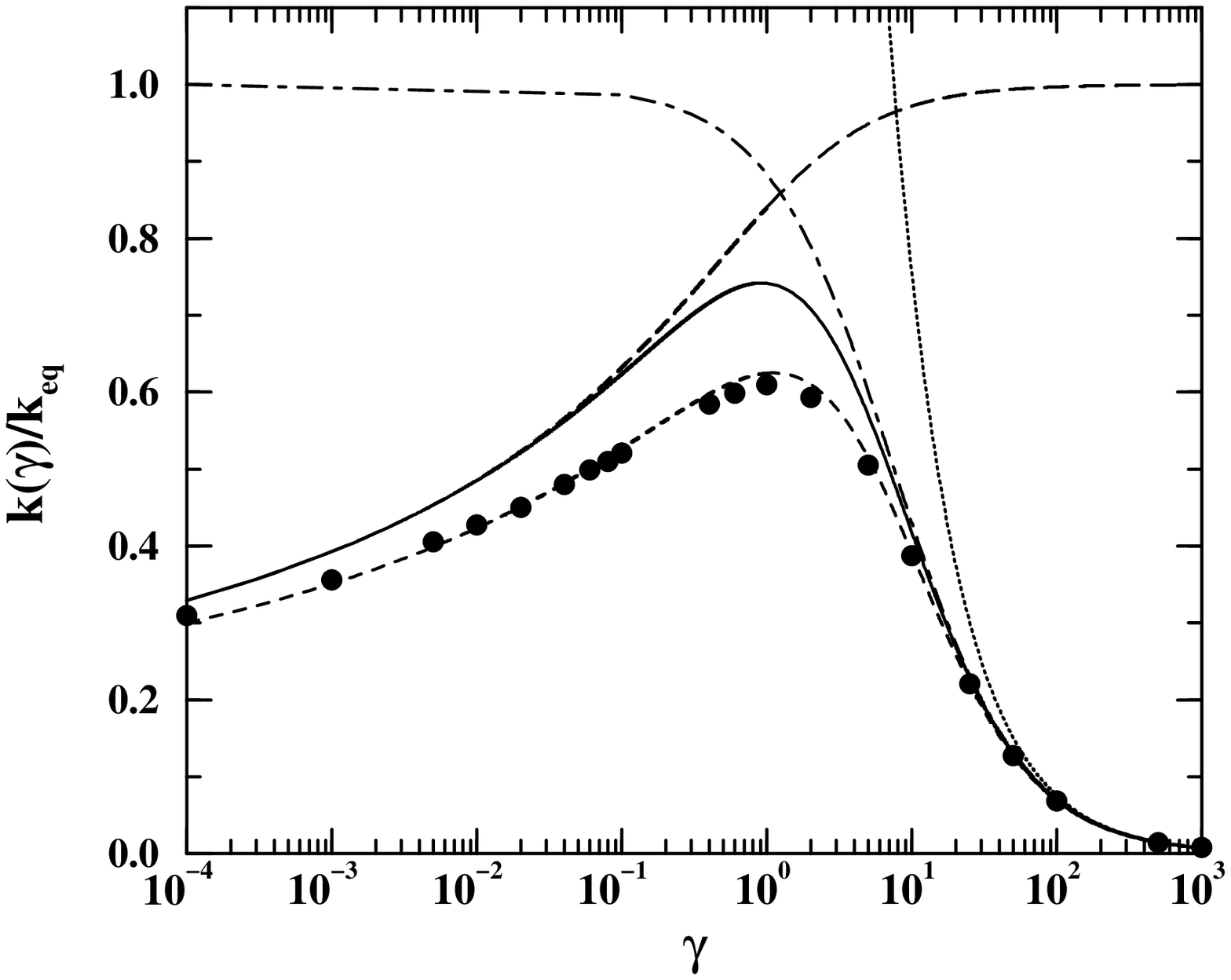,width=3.7in,angle=0}}
\vspace{0.25cm}
\caption{Reduced effective rate constant, $k/k_{\rm eq}$, as a function 
of the  friction coefficient $\gamma$. The data (closed circles) 
are obtained from Langevin dynamics simulations, the dot-dashed line 
corresponds 
to $k_{\rm high}$ given in Eq.(\ref{khigh}), the long-dashed line to 
$k_{\rm low}$ given in Eq.(\ref{klow}), the solid and dashed lines 
through the data represent the VMM interpolation in Eq.(\ref{mmk}) with 
$A=1.45$ and $A=6$, respectively. The dotted line is $k_{\rm d}/k_{\rm eq}$, 
where $k_{\rm d}$, given in Eq.(\ref{kd}), represents the purely 
diffusive part (i.e., $\kappa\rightarrow \infty$) of $k_{\rm high}$.}
\label{fig2}
\end{figure}

\vspace{0.25cm}

\begin{figure}[ht]
\vspace{0.3cm}
\centerline{\psfig{figure=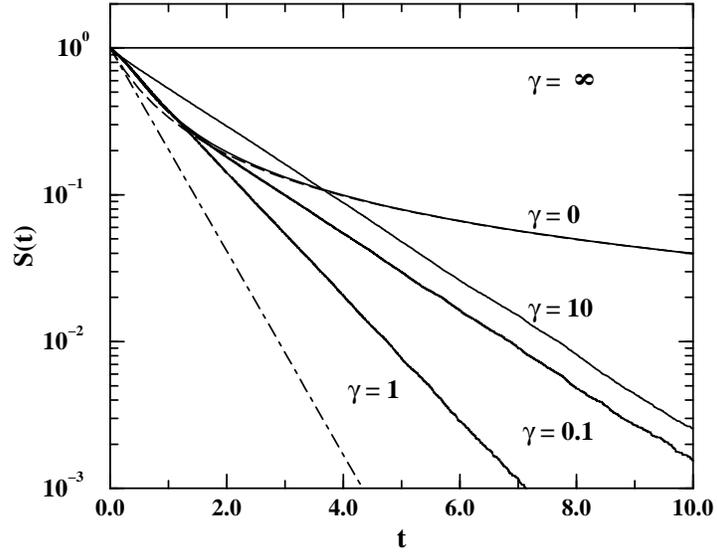,width=3.7in,angle=0}}
\vspace{0.25cm}
\caption{Survival probability $S(t)$ as a function of time for various 
values of $\gamma$. Solid lines correspond to the Langevin dynamics 
simulations results. For $\gamma=0$, the solid line represents the exact 
result in Eq.(\ref{sexact}) and the dashed line $S_{\rm low}(t)$ given in 
Eq.(\ref{slow}). The dot-dashed line is, ${\rm e}^{-2\kappa\,t}$.}
\label{fig3}
\end{figure}

\vspace{0.5cm}

\begin{figure}[ht]
\vspace{0.3cm}
\centerline{\psfig{figure=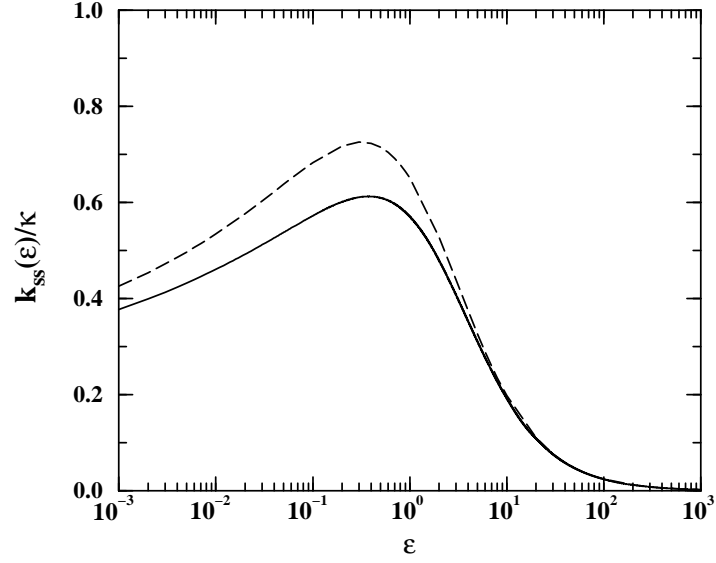,width=3.7in,angle=0}}
\vspace{0.25cm}
\caption{Reduced steady-state rate constant, 
$k_{\rm ss}(\varepsilon)/\kappa$, as a function of the dimensionless 
friction coefficient $\varepsilon=\gamma/c$. Solid line (Eq.(\ref{kss}) 
with $A=6$) is the essentially exact result while the dashed line 
(Eq.(\ref{kss}) with $A=1.45$) is obtained using the approximate theory.}
\label{fig4}
\end{figure}

\vspace{0.5cm}

\begin{figure}[ht]
\vspace{0.3cm}
\centerline{\psfig{figure=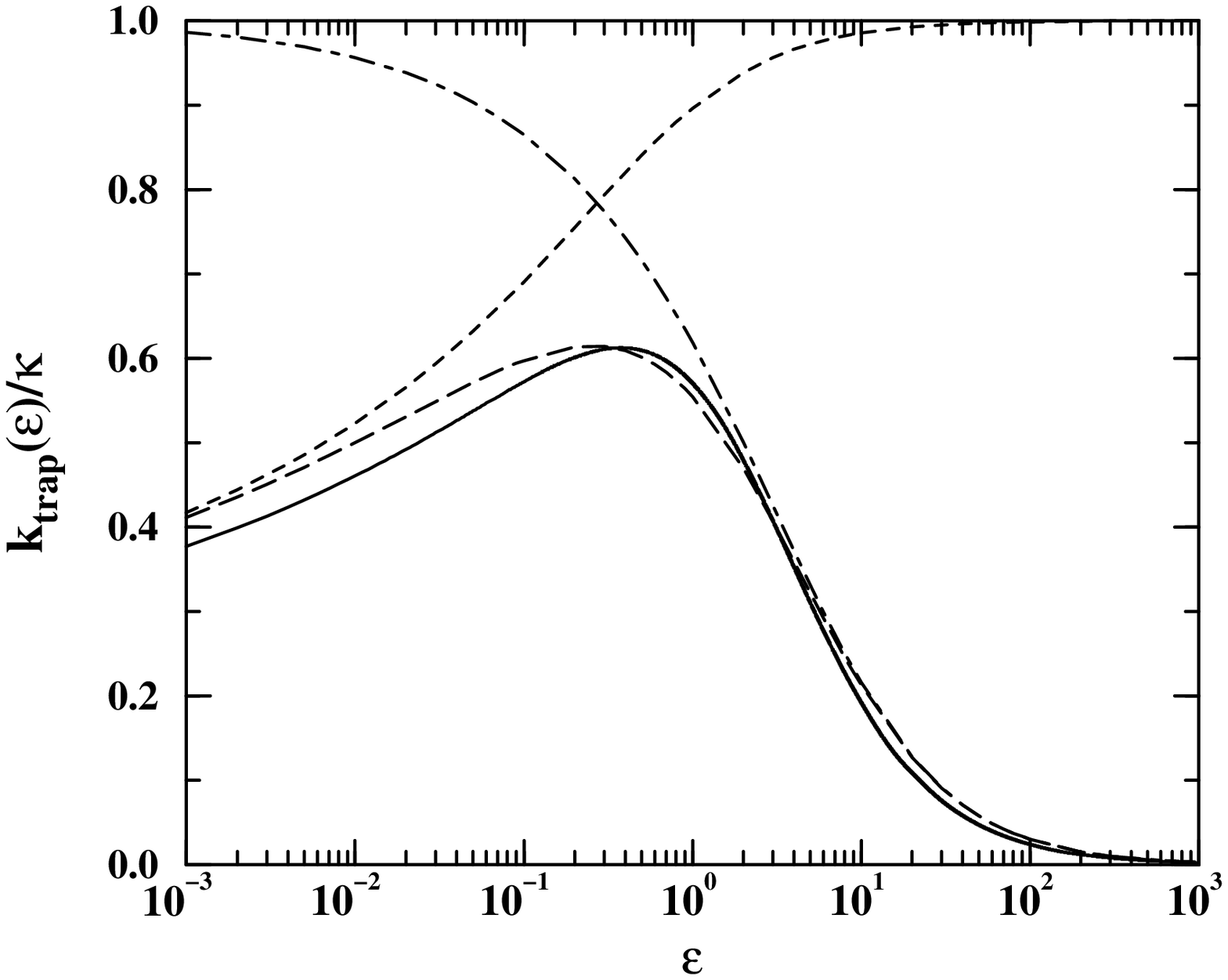,width=3.7in,angle=0}}
\vspace{0.25cm}
\caption{One-dimensional reduced steady-state rate constant, 
$k_{\rm trap}(\varepsilon)/\kappa$, for the trapping problem as a function 
of the dimensionless friction coefficient $\varepsilon=\gamma/c$. The solid 
line represents the exact result (Eq.(\ref{kss}) with $A=6$), the 
dot-dashed line corresponds to $k_{\rm targ}(\varepsilon)/\kappa$ given in 
Eq.(\ref{ksshigh}) for the target problem, the dashed line to 
$k_{\rm low}(\varepsilon)/\kappa$ given in Eq.(\ref{ksslow}) with $A=1.45$, 
and the long-dashed line represents the VMM interpolation in 
Eq.(\ref{ktrap}).}
\label{fig5}
\end{figure}

\vspace{0.5cm}

\begin{figure}[ht]
\vspace{0.3cm}
\centerline{\psfig{figure=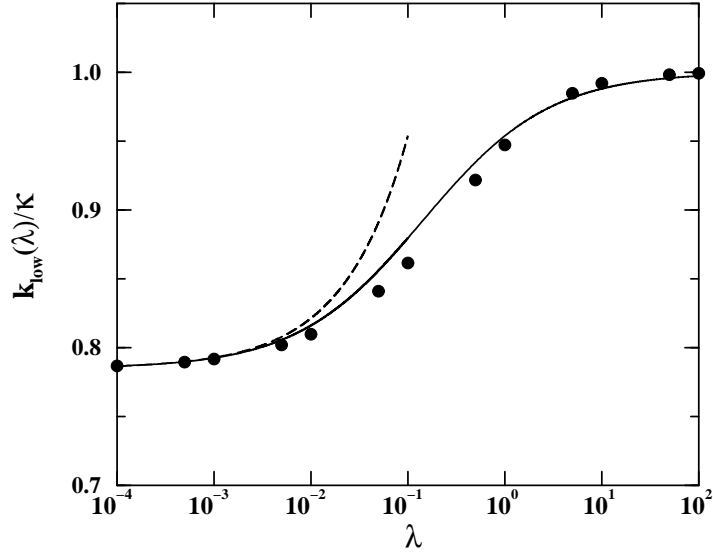,width=3.7in,angle=0}}
\vspace{0.25cm}
\caption{Three-dimensional reduced rate constant, $k_{\rm low}(\lambda)/\kappa$, 
in the low friction regime as a function of the parameter 
$\lambda=2\pi \varepsilon/3\phi$. The closed circles are obtained from 
numerical solution of Eq.(\ref{dtauv4}) , the solid line corresponds to 
expression in Eq.(\ref{ksslow3}) and the dashed line to the small-$\lambda$ 
behavior in Eq.(\ref{kll}).}
\label{fig6}
\end{figure}

\vspace{0.5cm}

\begin{figure}[ht]
\vspace{0.3cm}
\centerline{\hbox{
\hspace{-0.2cm}
\psfig{figure=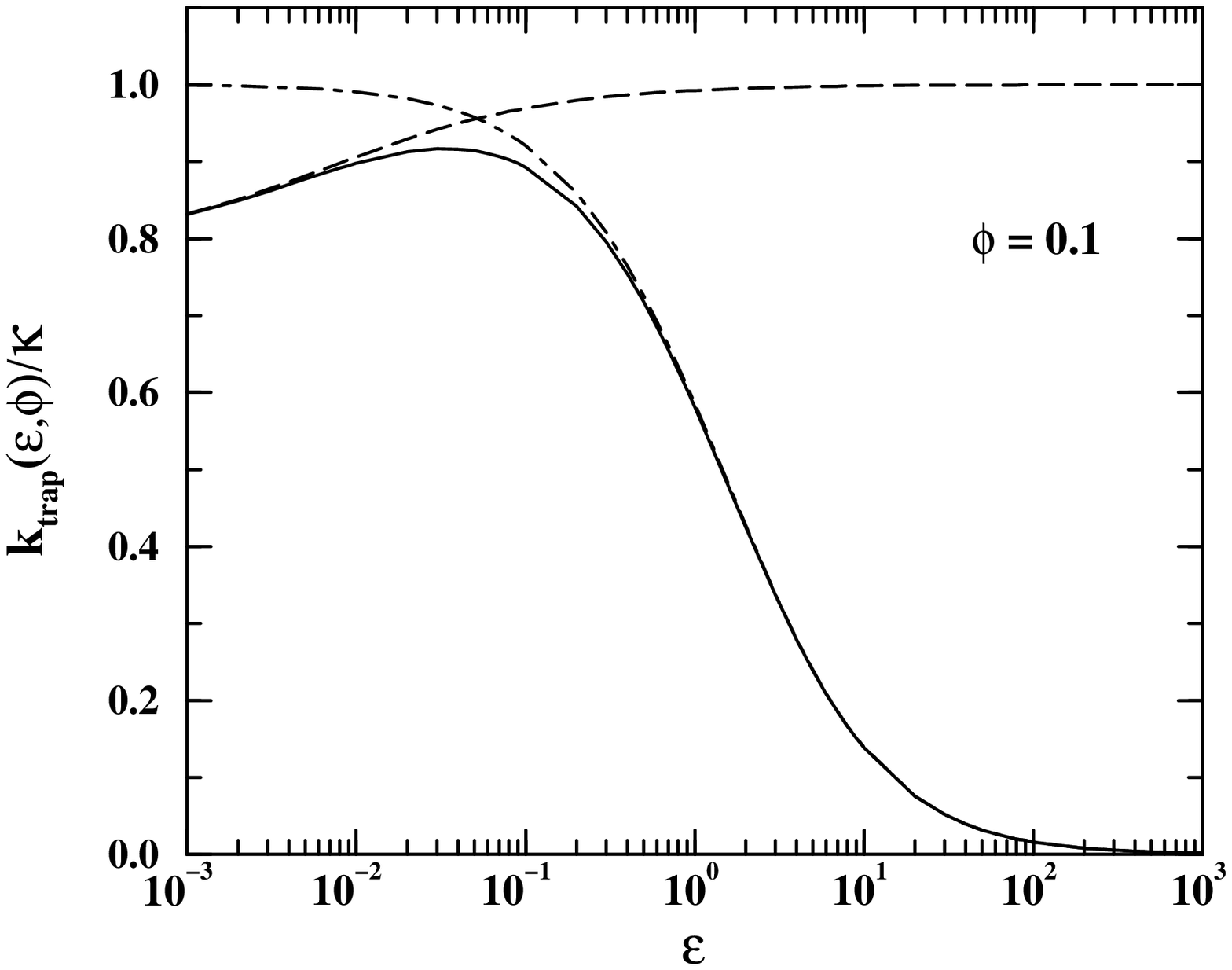,width=3.2in,angle=0}
\hspace{0.3cm}
\psfig{figure=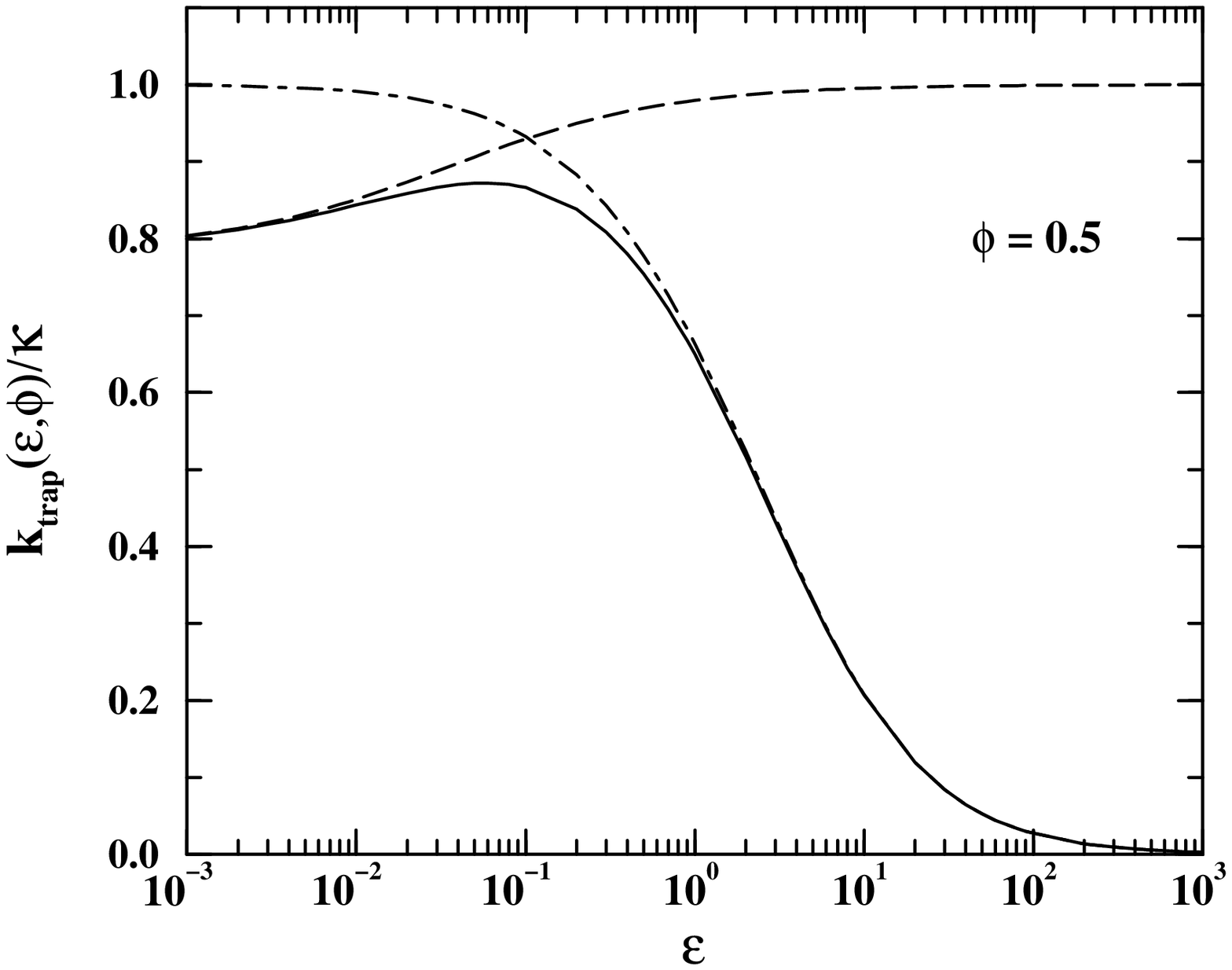,width=3.2in,angle=0}
}}
\vspace{0.25cm}
\caption{Three-dimensional reduced steady-state rate constant, 
$k_{\rm trap}(\varepsilon,\phi)/\kappa$, for the trapping problem as a function 
of the dimensionless friction coefficient $\varepsilon=\gamma R/\sqrt{2\pi}$ 
for two values of the volume fraction of traps $\phi$. The dot-dashed line 
corresponds to $k_{\rm targ}(\varepsilon,\phi)/\kappa$ given in 
Eq.(\ref{ktargck1}) for the target problem, the dashed line to $k_{\rm low}$ 
given in Eq.(\ref{ksslow3}), and the solid line represents the VMM 
interpolation in Eq.(\ref{mmktrap3}).}
\label{fig7}
\end{figure}


\newpage

\begin{thebibliography}{99}

\bibitem{Rice} S. A. Rice, {\it Diffusion-Limited Reactions}, (Elsevier, 
Amsterdam, 1985); E. Kotomin and V. Kuzovkov, {\it Modern Aspects of 
Diffusion-Controlled Reactions}, (Elsevier, Amsterdam, 1996).

\bibitem{Kramers}  H. A. Kramers, Physica \textbf{7}, 284 (1940); P. H\"{a}nggi,
P. Talkner and M. Borkovec, Rev. Mod. Phys. \textbf{62}, 251 (1990).

\bibitem{Melnikov} V. I. Mel'nikov and S. V. Meshkov, J. Chem. Phys. {\bf 85}, 
1018 (1986); V. I. Mel'nikov, Phys. Rep. \textbf{209}, 1 (1991).

\bibitem{Marsh} T. W. Marshall and E. J. Watson, J. Pys. A: Math. Gen. 
 \textbf{18}, 3531 (1985); {\it ibid.} \textbf{20}, 1345 (1987).

\bibitem{Wang}  M.S. Wang and G.E. Uhlenbeck, Rev. Mod. Phys. \textbf{17},
323 (1945).

\bibitem{Doering}  C.R. Doering in \textit{Unsolved Problems of Noise,} 
ed. C.R.Doering, L.B. Kiss and M.F. Shlesinger, (World Scientific, 
Singapore, 1997) p. 11 ; J. Masoliver and J.M. Porr\`{a}, 
Phys. Rev. E \textbf{53}, 2243 (1996).

\bibitem{Bicout} D. J. Bicout, A. M. Berezhkovskii, A. Szabo and 
G. H. Weiss, Phys. Rev. Lett. \textbf{83}, 1279 (1999). 

\bibitem{Visscher} P. B. Visscher, Phys. Rev. B \textbf{13}, 3272 (1976)

\bibitem{Stegun} M. A. Abramowitz and I. A Stegun, {\it Handbook of 
Mathematical Functions}, (Dover Publications Inc., New York, 1972).

\bibitem{Bicout1} D. J. Bicout and A. Szabo, 
J. Chem. Phys. \textbf{106}, 10292 (1997).

\bibitem{Bala} B. Va. Balagurov and V. G. Vaks, Zh. Eksp. Teor. Fiz. 
\textbf{65}, 1939 (1973) [Sov. Phys. JETP \textbf{38} 968 (1974)].

\bibitem{Szabo} A. Szabo, J. Phys. Chem.  \textbf{93}, 6929 (1989).

\bibitem{Harris} S. Harris, J. Phys. Chem.  \textbf{78}, 4698 (1983).


\end{thebibliography}
\end{document}